\documentclass[preprintnumbers,amsmath,amssymb,nofootinbib,prd]{revtex4-2}
\usepackage{afterpage}
\usepackage{subfigure}
\usepackage{rotating} 
\usepackage{ulem}
\usepackage{booktabs}
\usepackage{graphicx}
\usepackage{dcolumn}
\usepackage{bm}
\usepackage{float}
\usepackage{amsmath}
\usepackage{natbib}
\usepackage{hyperref}
\renewcommand{\arraystretch}{1.5}
\usepackage{xcolor}

\def\Wedge#1{\wedge^{#1}}

\newcommand{\beq}{\begin{equation}}
\newcommand{\eeq}{\end{equation}}
\newcommand{\bea}{\begin{eqnarray}}
\newcommand{\eea}{\end{eqnarray}}
\newcommand{\bseq}{\begin{subequations}}
\newcommand{\eseq}{\end{subequations}}

\newcommand{\hf}{\frac{1}{2}}
\newcommand{\al}{\alpha}
\newcommand{\be}{\beta}

\newcommand{\si}{\sigma}
\newcommand{\mn}{{\mu\nu}}

\def\prt{\partial}
\newcommand{\ie}{{\it i.e.}}

\newcommand{\bit}{\begin{itemize}}
\newcommand{\eit}{\end{itemize}}

\newcommand{\reply}[1]{\textcolor{black}{ #1}}

\hypersetup{
    colorlinks = true,
    allcolors = blue,
}

\begin{document}

\title{Constraints on birefringence-free photon theory within standard-model extension}
\author{Zhi Xiao}
\email{spacecraft@pku.edu.cn}
\affiliation{Department of Mathematics and Physics, North China Electric Power University, Beijing 102206, China}
\affiliation{Hebei Key Laboratory of Physics and Energy Technology, \\ North China Electric Power University, Baoding 071000, China}

\author{Hanlin Song}
\email{hsongphy@gmail.com}
\affiliation{School of Physics, Peking University, Beijing 100871, China}

\author{Bo-Qiang Ma}\email{mabq@pku.edu.cn}
\affiliation{School of Physics, Zhengzhou University, Zhengzhou 450001, China}
\affiliation{School of Physics, Peking University, Beijing 100871, China}

\begin{abstract}
Constraints on the birefringence-free subset of Lorentz-violating (LV) operators are derived using 14 GRB photons in the GeV-band. 
These constraints target the isotropic $c_{(I)00}^{(d)}$ coefficients for dimensions $d=6,8,10$ within the framework of the Standard-Model Extension (SME). 
Employing theory-agnostic Bayesian parameter estimation methods, our analysis indicates a preference for subluminal LV effects.
Focusing on this case,
we further refine the parameter constraints, yielding results that are mutually consistent. Within the 95\% posterior credible interval,
our constraints yield the bounds, 
$|c_{(I)00}^{(6)}|\le7.75 \times 10^{-20} ~ {\rm GeV}^{-2}$, $|c_{(I)00}^{(8)}|\le4.92 \times 10^{-24} ~ {\rm GeV}^{-4}$, and $|c_{(I)00}^{(10)}|\le3.46 \times 10^{-28} ~ {\rm GeV}^{-6}$,
which improve upon the most stringent credible-interval  bounds reported in the literature by at least five orders of magnitude.
\end{abstract}
\maketitle

\section{Introduction}
The Lorentz invariance, a cornerstone of special relativity, was first established within electrodynamics
and intrinsically tied to the constancy of speed of light {\it in vacuo}.
Intriguingly, the quest for a theory of quantum gravity suggests that the microstructure of vacuum may,
once again, be probed by studying light propagation.
A speculative yet insightful possibility is that the vacuum behaves as a kind of nontrivial optical medium,
a property that would inevitably break Lorentz invariance.

This concept of Lorentz-violating electrodynamics \cite{LVQED2002} has been substantially developed within the standard model extension (SME) \cite{SMEa, SME98},
an effective field theory (EFT) framework pioneered by Colladay, Kosteleck\'y, Mewes, 
and their collaborators.
The SME systematically encodes potential Lorentz violation (LV) into a set of tensor coefficients, thereby providing a comprehensive framework for testing Lorentz symmetry.
Beyond the SME \cite{SBLV1989a, SBLV1989b, SMEa, SME98}, several other theoretical frameworks have been proposed to test Lorentz violation, including space-time foam in string theory \cite{Amelino-Camelia:1996bln, Amelino-Camelia:1997ieq, Ellis:1999rz, Ellis:1999uh, Ellis:2008gg, Li:2009tt, Li:2021gah, Li:2021eza, Li:2024crc, Li:2025wqc, Li:2025uwn}, 
loop quantum gravity \cite{Gambini:1998it, Alfaro:1999wd,Li:2022szn}, doubly-special relativity (DSR) \cite{Amelino-Camelia:2002cqb, Amelino-Camelia:2002uql, Amelino-Camelia:2000stu, DSR2010}, weighted scale-invariant quantum field theories \cite{Anselmi2008}, and Hořava–Lifshitz theories \cite{Horava2009, AZW2017}. 

On the other hand, in the kinematic test of Lorentz symmetry using both photon and gravitational waves, 
it is possible to parameterize experimental results in a model-independent way \cite{MITest, Ellis:2005, Shao:2009bv, Zhang:2014wpb, Xu:2016zxi, Xu:2016zsa, Huang:2019etr, Zhu:2021pml, Zhu:2021wtw, Zhu:2022usw, Song:2024and, Song:2025qej, Song:2025myx, Song:2025akr, Song:2025ksi}. 
\reply{In particular, significant progress has been made in recent years in using high-energy GRB photons to probe LV \cite{FermiGBMLAT:2009nfe, Vasileiou:2013vra, Wei:2016exb, Ellis:2018lca, MAGIC:2020egb, Wei:2022zqi, LHAASO:2024lub, Guerrero:2025awe}.  For recent reviews, see, {\it e.g.} \cite{Addazi:2021xuf, He:2022gyk,AlvesBatista:2023wqm, Li:2025yvq}.}
In such searches, the key observable is the time delay 
during particle propagation,
together with its dependence on frequency or energy.
In general, this delay can be helicity dependent, giving rise to two distinct branches: superluminal and subluminal propagation. 
This phenomenon is called the vacuum birefringence.
However, photon birefringence is subject to extremely stringent constraints from polarization observations across different sources,
including cosmic microwave background (CMB) \cite{CMBLV-N,CMBLV-K1, CMBLV-K2, CMBLV-K3}, astrophysical objects \cite{Shao:2011uc}, active galactic nuclei (AGN) and gamma-ray bursts (GRB) \cite{Astro-K1, Polarimetri-GRB2006, CLV-GRB2013, Astro-K2, Astro-K3}.

To circumvent these limitations, one may restrict attention to a subset of Lorentz-violating parameters that do not induce vacuum birefringence at leading order.
Interestingly, this subset typically leads only to subluminal, frequency-dependent arrival time delay, thereby avoiding potential instability issues associated with the superluminal branch.
The superluminal photon scenario is tightly constrained by the possibility of photon decay into lepton pairs
$\gamma\rightarrow e^+\,+\,e^-$
or photon spliting $\gamma\rightarrow3\gamma$ \cite{Threshold03, AK03gamma}. 
In this work, we focus on the birefringence-free subset of the photon sector of SME 
and using high-energy GRB 
photons
to place constraints.
It is well established that the LV effects in photon propagation can manifest as birefringence, dispersion, and anisotropy 
\cite{LVQED2002,LVQED2009}.
The birefringence-free subset remains capable of producing both dispersive and anisotropic signatures.

As we focus on the high-energy photons 
in the GeV range, it becomes more appropriate to constrain operators with mass dimension greater than 
four
--- the so-called irrelevant operators.
From the perspective of 
EFT, one expects that constraints derived from GeV  $\gamma$-ray photons should be significantly stronger
for power-counting non-renormalizable LV operators than those obtained with lower-energy photons ({\it e.g.} Ref. \cite{JJWei2017}). As we will show, our calculations indeed confirm this expectation. 
The only limitation is that its statistical significance, or credible level, 
is much lower than that of the constraints derived from low-energy $\gamma$-ray photons. 
This is primarily because low-energy $\gamma$-ray events are far more abundant, 
and such photons are comparatively easier to detect. 
However, with the continued accumulation of high-energy $\gamma$-ray events, the statistical significance can ultimately be improved. 
In this work, we emphasize the advantages of GeV $\gamma$-ray photons in probing LV physics. 

The remainder of this work is organized as follows. 
In Sec. \ref{BRofSME}, we will briefly review the photon sector of the SME, 
primarily focusing on the birefringence-free subset and its parameterization in both Cartesian coordinates and spherical harmonics.
In Sec. \ref{DAHS}, using this parameterization and the GRB observational data, particularly  photon events with energies above the GeV scale,
we derive bounds on the birefringence-free LV coefficients for operators with dimension-6, -8, -10 based on arrival-time delay.
Finally, in Sec. \ref{results}, we provide discussions and future perspectives.


\section{Photon sector in SME}\label{BRofSME}
The action of the 
LV
Maxwell theory \cite{LVQED2002,LVQED2009} in SME is
\bea\label{Maxwell-Full}&&
\hspace{-6mm}
I_{A}=\int\sqrt{-g}\,d^4x\left[\mathcal{L}_{0}
+\delta\mathcal{L}_\mathrm{LV}\right],
\eea
where $\mathcal{L}_{0}$ is composed of the conventional Lorentz invariant Maxwell term and the current interaction term
\bea&&\label{Photon-LI-d}
\mathcal{L}_\mathrm{0}\equiv-\frac{1}{4}g^{\mu\rho}g^{\nu\si}F_{\mn}F_{\rho\si}+j_e^\mu A_\mu,
\eea
and the LV photon Lagrangian $\delta\mathcal{L}_\mathrm{LV}=\Delta\mathcal{L}_\mathrm{e}+\Delta\mathcal{L}_\mathrm{o}$ can be separated into CPT-even and CPT-odd parts,
$\Delta\mathcal{L}_\mathrm{e}$ and $\Delta\mathcal{L}_\mathrm{o}$, respectively,
distinguished by their transformation properties under the combined discrete symmetries of charge conjugation, parity reversal, and time inversion (collectively abbreviated as CPT). 
The two parts are superficially similar in form (by just look at the notations) to those in the minimal SME,
\bea&&\label{Photon-LV-d}
\Delta\mathcal{L}_\mathrm{e}=-\frac{1}{4}(\hat{k}_F)^{\mn\rho\si}F_{\mu\nu}F_{\rho\si},\qquad~~~~~~~~
\Delta\mathcal{L}_\mathrm{o}=\hat{k}_\mu({^*F}^{\mn})A_\nu
\eea
where ${^*F}^{\mn}=\frac{1}{2}\epsilon^{\mn\al\beta}F_{\al\beta}$ is the dual Faraday tensor.
For notational simplicity, we set $\hat{k}_\mu=(\hat{k}_{AF})_\mu$, 
and in general,
\bea\label{kAF&kF}
\hat{k}_\mu\equiv\sum_{d=\mathrm{odd}}(k^{(d)})_{\mu}^{~{\nu_1}...{\nu_{d-3}}}\prt_{\nu_1}...\prt_{\nu_{d-3}},
\qquad
(\hat{k}_F)^{\mn\rho\si}\equiv\sum_{d=\mathrm{even}}(k_F^{(d)})^{\mn\rho\si{\nu_1}...{\nu_{d-4}}}\prt_{\nu_1}...\prt_{\nu_{d-4}}.
\eea
Note the LV operators in (\ref{Photon-LV-d}) are built from two sets of LV coefficients,
$(k^{(d)})_{\mu}^{~{\nu_1}...{\nu_{d-3}}}$ and $(k_F^{(d)})^{\mn\rho\si{\nu_1}...{\nu_{d-4}}}$,
a set of partial derivatives and photon fields.
The above quadratic photon Lagrangian can lead to equation of motion \cite{LVQED2009},
\bea\label{photon-eom}
\prt_\nu\,G^{\mn}\equiv\prt_\nu[F^{\mn}+(\hat{k}_F)^{\mn\rho\si}F_{\rho\si}-2\epsilon^{\mn\rho\si}\hat{k}_\rho\,A_\si]=j_e^\mu.
\eea
It is easy to check that the current conservation $\prt_\mu\,j_e^\mu=0$ is guaranteed by the antisymmetry of $G^{\mn}=-G^{\nu\mu}$. 
Apart from the third term involving $\hat{k}_\rho$, the above expression is explicitly gauge invariant.
The term involving $\hat{k}_\rho$ is also gauge invariant provided that $\prt_{[\nu}\hat{k}_{\rho]}=0$,
a condition naturally satisfied when the tensor coefficients are taken to be constant in flat spacetime in Cartesian coordinates. 
Gauge invariance follows automatically from $\prt_\nu[\epsilon^{\mn\rho\si}\hat{k}_\rho\,A_\si]=\hf\epsilon^{\mn\rho\si}\hat{k}_\rho\,F_{\nu\si}$.

Adopting the plane wave ansatz $A_\mu(x)=A_\mu(p)e^{-p\cdot x}$ and far from the source tube such that $j_e^\mu=0$,
Eq. (\ref{photon-eom}) can be reduced to the matrix equation
$\underline{M}^{\mn}A_\nu=0$ with
\bea&&\label{Matrix-eom}
\underline{M}^{\mn}=2\hat{\chi}^{\mu\rho\nu\si}p_\rho p_\si+2i\hat{X}^{\mn\rho}p_\rho,\quad
\hat{\chi}^{\mu\rho\nu\si}
=[\eta^{\mn}\eta^{\rho\si}-\eta^{\mu\rho}\eta^{\nu\si}+2(\hat{k}_F)^{\mu\rho\nu\si}],\quad
\hat{X}^{\mn\rho}=-\epsilon^{\mn\rho\si}\hat{k}_\si.
\eea
The underline of $\underline{M}^{\mn}$ indicates that the matrix is in momentum space, where all the partial derivatives $i\prt_{\mu_j}$ in $(\hat{k}_F)^{\mn\rho\si}$ and $\hat{k}_\rho$ have been replaced by $p_{\mu_j}$.
Then one can quickly check that $p_\mu\underline{M}^{\mn}=0=\underline{M}^{\mn}p_\nu$,
where the former is due to charge conservation and the latter is due to gauge invariance.
The gauge invariance also indicates that $\underline{M}^{\mn}$ is a singular matrix with $\mathrm{det}[\underline{M}]=0$.
To have non-trivial physical solutions, we can impose a gauge condition, 
such as Coulomb gauge $p^a\cdot A_a=0$ (where $a$ ranges over spatial indices $1,2,3$)
or radiation gauge $A^r=0$, to get a reduced matrix $M$ from Eq. (\ref{Matrix-eom}).
Then the reduced matrix is not singular in most situations, 
and from the matrix equation $\mathrm{det}[{M}]=0$, we can get physical dispersion relations \cite{XM-photon2009}.
However, imposing gauge conditions generally breaks observer Lorentz covariance,
which not only compromises aesthetic elegance, 
but also makes the resulting dispersion relations appear gauge dependent.

Lucky that there is a covariant way to obtain dispersion relations,
which largely utilize {\it the linear independence of vector space and the nullity of the square matrix $\underline{M}^{\mn}$} in linear algebra \cite{Itin-Cdr2009,LVQED2009}.
As introduced in Ref. \cite{LVQED2009}, a very elegant way to express the linear independence is to use the wedge product to define the multilinear map
$(\Wedge{n}\underline{M})$ \cite{LVQED2009}, which is a generalization of the linear map $\underline{M}^{A}_{~B}\chi^B=\xi^A$
between the two vector spaces spanned by $\{\chi^A,~A=0,1,...N-1\}$ and $\{\xi^A,~A=0,1,...N-1\}$,
where $N$ is the spacetime dimension.
The multilinear map defines a map between two $n-$vectors, $\omega_A\equiv A^{a_1}\wedge\,A^{a_2}...\wedge A^{a_n}$
and $\omega_B\equiv (\wedge^n\underline{M})\omega_A=(\underline{M}A^{a_1})\wedge....\wedge (\underline{M}A^{a_n})$.
Note that an ``n-vector" can be regarded as an n-form constructed from n distinct 1-forms $\{A^{a_i},~i=1,...,n\}$, 
where the labels $a_i$ serve only as bookkeeping indices distinguishing the different 1-forms rather than as spacetime indices of a specific vector. 
Otherwise the notation $A^{a_i}$ may be mistakenly interpreted as a contravariant vector component.

The covariant dispersion relation can be expressed as the vanishing of the trace of the dual multiplicity-3 linear map,
$(\widetilde{\wedge^3\underline{M}})^\rho_{~\rho}=0$.
The underlying logic \cite{LVQED2009} follows from the rank-nullity theorem.
Since $\mathrm{det}[\underline{M}]=0$, the gauge-invariant matrix $\underline{M}$ is singular,
and the condition $\underline{M}^{\mn}p_\nu=0$ forces the linear space on which $\underline{M}^{\mn}$ acts to admit a gauge mode,
further reducing its rank. Therefore $\mathrm{rank}[\underline{M}]\le2$.
This can be equivalently expressed by the exterior-power condition, $\wedge^3\underline{M}=0$.
In four-dimensional spacetime, the dual of the multiplicity-3 linear map is one-dimensional,
so $(\widetilde{\wedge^3\underline{M}})$ is intrinsically a $4\times4$ matrix.
With $\underline{M}^{\mn}p_\nu=0$ and $\underline{M}^\dagger=\underline{M}$,
$\wedge^3\underline{M}=0$ can be further expressed as 
$(\widetilde{\wedge^3\underline{M}})^\rho_{~\rho}=0$,
which corresponds to a scalar covariant dispersion relation.
Non-birefringence requires degeneracy between the two polarizations, 
which imposes the stronger condition $\wedge^2\underline{M}=0$,
corresponding to a even lower rank of $\underline{M}$ \cite{LVQED2009}.

It is clear that the CPT-odd $\hat{k}_\al$ coefficients necessarily induce birefringence:
by breaking Parity symmetry, they cause left- and right- circular polarizations to propagate differently,
as shown in Ref. \cite{CFJ1990} and in axion electrodynamics \cite{axionEM}.
Therefore, in the next subsection we omit the $\hat{k}_\al$ coefficients.

\subsection{Birefringence-free subspace}
As our primary concern is the birefringence-free subspace, 
we may adopt the notation of Ref. \cite{LVQED2009} to decompose 
\bea\label{Riemann-decomp}
(\hat{k}_F)_{\mn\rho\si}=\frac{2}{N-2}[g_{\mu[\rho}(\hat{c}_F)_{\si]\nu}
-g_{\nu[\rho}(\hat{c}_F)_{\si]\mu}]+\hat{C}_{\mn\rho\si},
\eea
where the Weyl-like coefficient $\hat{C}_{\mn\rho\si}$ is the non-metric part 
and thus definitely leads to birefringence.
The absence of vacuum birefringence implying $\hat{C}_{\mn\rho\si}=0$, admits an alternative interpretation. In the framework of premetric electromagnetics, $(\hat{k}_F)^{\mn\rho\si}$ can be viewed as part of the constitutive tensor $\hat{\chi}^{\mn\rho\si}$,
which determines the entire optical behavior of electromagnetic waves.
In particular, it determines the inhomogeneous Maxwell equation $\prt_\mu H^{\mn}=j^\nu$ through the induced electromagnetic excitation two-form
$H^{\mn}=\hf\hat{\chi}^{\mn\rho\si}F_{\rho\si}$.
In general, the corresponding wave equation is quartic, allowing vacuum birefringence.
However, as shown in Ref. \cite{premetric2005}, imposing the reciprocal and closure relation \cite{premetric2000}
induces duality symmetry, which provides a sufficient condition \cite{LVQED2009} for birefringence-free optics, and eliminates the non-metric part $\hat{C}_{\mn\rho\si}$ of the constitutive tensor.
In this case, the quartic dispersion relation reduces to a quadratic one, 
yielding a unique light cone.

On the other hand, it is more convenient to define a set of 3-dimensional matrices from $(\hat{k}_F)^{\mn\rho\si}$ \cite{LVQED2009} as follows,
\bea\label{kappa-e-o}
\hat{\kappa}_{e+}=\hf[\hat{\kappa}_{DE}+\hat{\kappa}_{HB}]-\hat{\kappa}_{\mathrm{tr}+},\quad
\hat{\kappa}_{e-}=\hf[\hat{\kappa}_{DE}-\hat{\kappa}_{HB}]-\hat{\kappa}_{\mathrm{tr}-},\quad
\hat{\kappa}_{o-}=\hf[\hat{\kappa}_{DB}-\hat{\kappa}_{HE}],\quad
\hat{\kappa}_{o+}=\hf[\hat{\kappa}_{DB}+\hat{\kappa}_{HE}],
\eea
where $\hat{\kappa}_{\mathrm{tr}\pm}\equiv\frac{1}{6}\mathrm{tr}[\hat{\kappa}_{DE}\pm\hat{\kappa}_{HB}]$ are two scalars and the definition of $\hat{\kappa}_{DE},~\hat{\kappa}_{HB},~\hat{\kappa}_{DB},~\hat{\kappa}_{HE}$
are as following
\bea
(\hat{\kappa}_{DE})^{ij}\equiv-2(\hat{k}_F)^{0i0j},\quad
(\hat{\kappa}_{HB})^{ij}\equiv\hf\epsilon^{imn}\epsilon^{jkl}(\hat{k}_F)^{mnkl},\quad
(\hat{\kappa}_{DB})^{ij}=-(\hat{\kappa}_{HE})^{ji}\equiv\epsilon^{jkl}(\hat{k}_F)^{0ikl}.
\eea

It is straightforward to verify that the first three matrices in Eqs.~(\ref{kappa-e-o})
are traceless and symmetric, \ie, $\hat{\kappa}_{e\pm}^T=\hat{\kappa}_{e\pm}$ and $\hat{\kappa}_{o-}^T=\hat{\kappa}_{o-}$.
The last matrix is antisymmetric, namely, $\hat{\kappa}_{o+}^T+\hat{\kappa}_{o+}=0$.
Although these definitions resemble those of the minimal SME \cite{LVQED2002}, they count only the free indices contracted with the Faraday tensor $F_{\mn}$.
Unlike in the minimal SME, both $\hat{\kappa}_\mathrm{tr+}$ and $\hat{\kappa}_\mathrm{tr-}$ can contribute to LV photon propagation \cite{LVQED2009}.

In the vacuum case, as $\hat{\kappa}_{e+}$ and $\hat{\kappa}_{o-}$ are related to Weyl-like tensor, they do contribute to birefringence.
The other remain 8 coefficients $\hat{\kappa}_{e-}$ and $\hat{\kappa}_{o+}$ and the 2 trace terms $\hat{\kappa}_{\mathrm{tr}\pm}$ are
the birefringence-free part. In fact, these 10 terms can be written as the metric-like $(\hat{c}_F)$ coefficients as below,
\bea
(\hat{c}_F)^{00}=\hf(3\hat{\kappa}_{\mathrm{tr}-}+\hat{\kappa}_{\mathrm{tr}+}),\quad
(\hat{c}_F)^{0i}=-\hf\epsilon^{ijk}(\hat{\kappa}_{o+})^{jk},\quad
(\hat{c}_F)^{ij}=\hf[\hat{\kappa}_{\mathrm{tr}-}-\hat{\kappa}_{\mathrm{tr}+}]\delta^{ij}-(\hat{\kappa}_{e-})^{ij}.
\eea
These 10 terms are the main concern of our work.
Then the effective Lagrangian $\Delta\mathcal{L}_e$ reduces to
\bea
\Delta\mathcal{L}_e\supset-\frac{1}{4}[(\hat{c}_F)^{\lambda\nu}F^\mu_{~\lambda}-(\hat{c}_F)^{\kappa\mu}F_{~\kappa}^{\nu}]F_{\mn},
\eea
from which we can easily deduce the dispersion relation at leading order of 
the $\hat{c}_F$ coefficients as below \cite{LVQED2009},
\bea
\omega^2=|\vec{p}|^2+p\cdot \hat{c}_F\cdot p\Rightarrow \omega\simeq (1+\varsigma^0) |\vec{p}|,\quad
v_g\equiv\frac{\prt\omega}{\prt|\vec{p}|}\simeq (1-\varsigma^0)\equiv
[1-\frac{1}{2\omega^2}(\hat{c}_F)^{\mu\nu}p_\mu p_\nu],
\eea
where $p\cdot \hat{c}_F\cdot p\equiv p^\mu(\hat{c}_F)^{\mu\nu}p_\nu$, $\varsigma^0\equiv p\cdot \hat{c}_F\cdot p/2\omega^2$, and an implicit assumption $\omega\simeq|\vec{p}|$ has been used at leading-order approximation for astrophysical photons.

\subsection{Birefringence-free terms in spherical harmornics}\label{SphericHarmornic}
To describe electromagnetic signals from distant astrophysical sources, it is highly effective to use spin-weighted spherical harmonics (SWSHs).
Unlike the spherical harmonics $Y_{lm}$, which form the basis eigenspace for the orbital angular momentum operator 
and can be used to decompose any scalar function into a sum of angular distributions with definite polar and azimuthal quantum numbers,
SWSHs are necessary for tensor functions. 
This is because there is no way to perform ordinary differential operations for tensor functions on a sphere;
instead, covariant differential operations have to be used on curved space. 
One way to avoid connections and covariant derivatives is to project tensor fields onto a set of null tetrads, 
which yields scalar functions that are spin-weighted.
The advantage of using SWSHs to expand tensor functions --- 
especially electromagnetic and gravitational fields --- is that 
both the amplitude and the polarization (or helicity, in the case of massless particles) are physically relevant for describing the observed waves. Strictly speaking, SWSHs must be defined on the fiber bundle over $S^3$ rather than the sphere $S^2$, 
although the latter is more intuitive and most commonly referenced in the literature. 
In fact, this subtlety arises from the inherent difference between scalar spherical harmonics and SWSHs: 
as complete basis functions, the former remain unchanged under rotations, 
while the latter transform. 
That is why SWSHs must be defined on both the base space $S^2$ and the tangent space --- collectively, the fiber space. 
Here, $S^2$ typically refers to the celestial sphere,
which is ideally suited for studying astrophysical electromagnetic or gravitational waves, 
as these waves carry information about their source and have an asymptotic boundary that is simply $S^2$. 
In precise geometric terms, the pole and flag structures introduced by Penrose and Rindler \cite{PRspinor} 
provide an intuitive framework for understanding the bundle space on which spin-weighted spherical harmonics are properly defined.

In practice, SWSHs are extensively used in astrophysics \cite{SWSHCMB97}, 
particularly for analyzing the polarization states of electromagnetic radiations in cosmic microwave background \cite{CMBSH} 
and recently observed gravitational waves \cite{GWSH1,GWSH2}.
The key advantage of SWSHs 
is their ability to provide rotation-invariant descriptions independent of the frame chosen in the tangent space of celestial sphere. 
Kosteleck\'y and Mewes innovatively applied SWSHs to describe more general tensor fields \cite{LVQED2009}, 
such as LV observer tensor fields.
These LV tensor fields, or their corresponding LV coefficients, transform as tensors under the observer Lorentz group 
but remain invariant (as dynamical scalars) under particle Lorentz transformations \cite{SMEa}. 
The primary benefit of decomposing LV coefficients using SWSHs is that 
these coefficients transform in a simple way under rotation about the light-of-sight direction.
This approach allows for a clear separation of the frequency and directional dependence of LV coefficients,
significantly simplify the analysis of associated LV effects. 

As for LV coefficients in quadratic electrodynamics, the CPT-odd $\hat{k}_{AF}$ coefficients necessarily induce vacuum birefringence 
while only a subset of the CPT-even $\hat{k}_{F}$ coefficients are birefringent-free in linear order. 
Strictly speaking, there are no birefringence-free quadratic photon LV coefficients to all orders \cite{Marco2015,Exirifard2010,ZX2025}.
From the analysis in Ref. \cite{LVQED2009}, 
only the subset of party even $E$-type coeffcients $(\hat{c}_F^{(d)})^{0E}_{njm}$ (corresponding to $\hat{c}_F$)
does not lead to vacuum birefringence and is our concern here. 
In $(\hat{c}_F^{(d)})^{0E}_{njm}$, the superscripts ``$d$" and ``$0E$" indicate mass dimension $d$, parity even and spin-weight $s=0$, respectively.
The superscript ``$0E$" may be understood from a reversal perspective, namely, birefringence.
A spin-weight of $s=0$ signifies no rotation in tangent frames, 
since birefringence-free propagation implies no precession of linear polarization in vacuum.
This arises from the absence of phase-velocity difference between left and right circular polarizations.
Moreover, parity violation is tied to birefringence,
as the symmetry between left and right circular polarization is lost.
Thus, it is expected that birefringence-free operators are parity even and spin-weight $0$.
The subscripts $n$ and $jm$ indicate frequency and angular (directional) dependence, respectively.
For gravitational waves, one may parameterize the corresponding LV coefficients in a similar way \cite{GWSH1}. 
 
As a side remark, note that within the SME framework, 
any linear correction to the dispersion relation must be birefringent, 
since the relevant operator has mass dimension 5 and originates from the $(\hat{k}^{(5)}_{AF})$ term. 
On the other hand, one can always introduce a birefringence-free linear correction to the dispersion relation in a model-independent way,
and such a correction would typically stem from some dimension-5 operator. 
However, this setup is not quite nature, 
since most admissible mass dimension-5 operators are axial 
in form and therefore breaks parity symmetry,
which in turn forces the dispersion relation to be helicity-dependent, especially when a vector or tensor background field 
is involved \cite{SMEa,SME98}. 
In loop gravity, a linear dispersion relation can remain helicity-independent only
by neglecting helicity-dependent terms, owning to stringent observational limits on polarization rotation \cite{SpeedVariation2023}.
Even in LV supersymmetric QED \cite{LVSUSYQED}, where a dimension-6 operator induces birefringent-free photon dispersion relation, the helicity degeneracy is protected by supersymmetry. 
Thus, we may conjecture that no  birefringence-free, LV dimension-5 operator exists within an EFT framework. 

As for GRB observations, there are two cases we can explore separately:
\begin{itemize}
    \item Isotropic case. In a special reference frame, we may have only isotropic part of $(\hat{c}_F)_{\mn}$ nonzero, \ie, $(c_F^{(d)})^{(0E)}_{n00}\neq0$. 
    In this case, as long as $d>4$, we are exploring the dispersive effect only.
    \item Anisotropic case.
    In general, all $(c_F^{(d)})^{(0E)}_{njm}\neq0$, and we are exploring not only the dispersive, but also the anisotropic effect.  
\end{itemize}
If data is abundant, we should explore both cases separately.
However, as previously mentioned, the high-energy $\gamma$-ray events are quite rare, 
and up to now, we have only accumulate a handful of events with known redshift. 
Due to the data limitations, unlike the situation of lower-energy $\gamma$-ray events, 
we study the isotropic case exclusively. In other words, we focus primarily on the $(c_F^{(d)})^{(0E)}_{n00}$ coefficients.

However, there is a notational 
distinction arises
when transitioning from the $c_F^{(d)}$-type coefficients, originating in the non-minimal SME \cite{LVQED2009,Koste-arbi13,Koste-arbi19}, to the birefringence-free $c_{(I)}^{(d)}$-type coefficients used in the analysis of arrival-time delays. 
The subscript ``I" for $c_{I}^{(d)}$ 
does not indicate isotropy,
rather, it denotes {\it intensity},
since the other coefficients $\vec{\varsigma}=(\varsigma^1,\varsigma^2,\varsigma^3)$ (see \cite{Kostelecky:2008be, LVQED2009}) are associated with the Stokes parameters $(Q,U,V)$.
Accordingly, the relevant LV coefficients are labeled as follows: 
\bea&&
\varsigma^0\sim c^{(d)}_{(I)jm},\quad \varsigma^3\sim k^{(d)}_{(V)jm}, \quad s=0\\&&
\varsigma^1\pm i\varsigma^2\sim  [k^{(d)}_{(E)jm}\mp i k^{(d)}_{(B)jm}],
\quad s=2.
\eea
The last relation with $s=2$ exactly matches the spin-weight of $Q\pm i\,U$ for photon fields. 
So it is no doubt that ``I" for $c_{I}^{(d)}$  is for intensity $I$ 
and thus has spin-weight 0. 
The two birefringence-free parameters are related by \cite{Kostelecky:2008be, LVQED2009}
{\small
\bea&&\label{Rel-cF-cI}
c_{(I)jm}^{(d)}\propto \sum_n(-1)^j
\left[(n+1)(n+2)(c_F^{(d)})^{0E}_{(n+2)\,jm}+(d-n-3)(d-n-2)(c_F^{(d)})^{0E}_{n\,jm}+2(d-n-3)(n+1)(c_F^{(d)})^{0E}_{(n+1)\,jm}\right],
\eea
}
\hspace{-1mm}The proportionality symbol ``$\propto$" indicates equality up to an overall numerical factor. However, in the astrophysical constraints expressed in terms of $c_{(I)jm}^{(d)},~k_{(V)jm}^{(d)},...$, the exact relation (\ref{Rel-cF-cI}) is not of major concern; we only need to recognize that a relation exists between $c_{(I)jm}^{(d)}$ and $(c_{F})_{n\,jm}^{(d)}$, and the sign factor $(-1)^j$ arises due to the fact that the direction along the line of sight $\hat{n}$ is opposite to the photon propagation direction $\hat{\vec{p}}$, \ie, $\hat{n}=-\hat{\vec{p}}$ \cite{LVQED2009}.

\begin{figure*}[htbp] 
	\centering 
    \begin{minipage}{0.33\textwidth}
        \centering
        \includegraphics[width=0.95\linewidth]{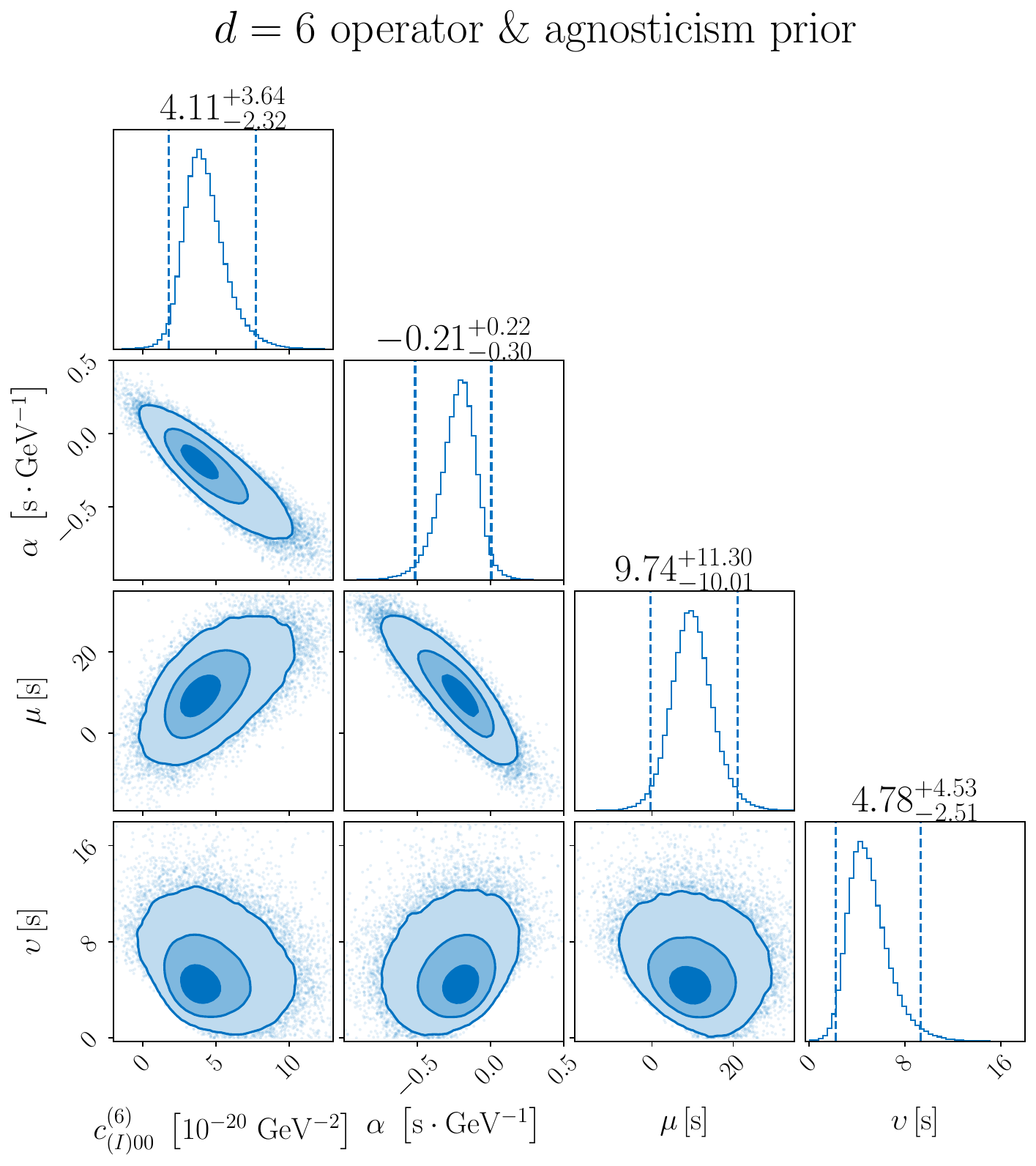}
    \end{minipage}\hfill
    \begin{minipage}{0.33\textwidth}
        \centering
        \includegraphics[width=0.95\linewidth]{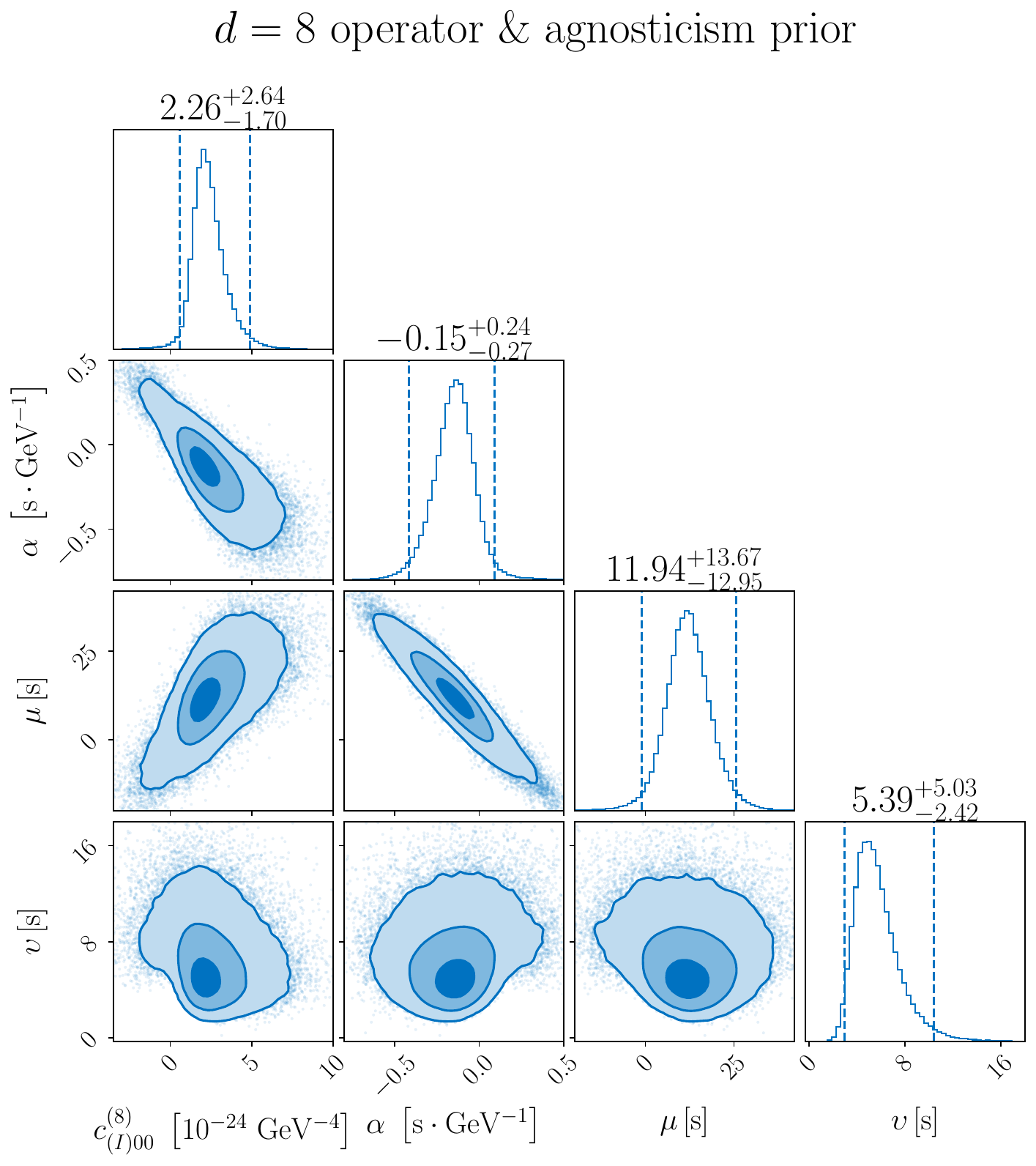} 
    \end{minipage}\hfill
    \begin{minipage}{0.33\textwidth}
        \centering
        \includegraphics[width=0.95\linewidth]{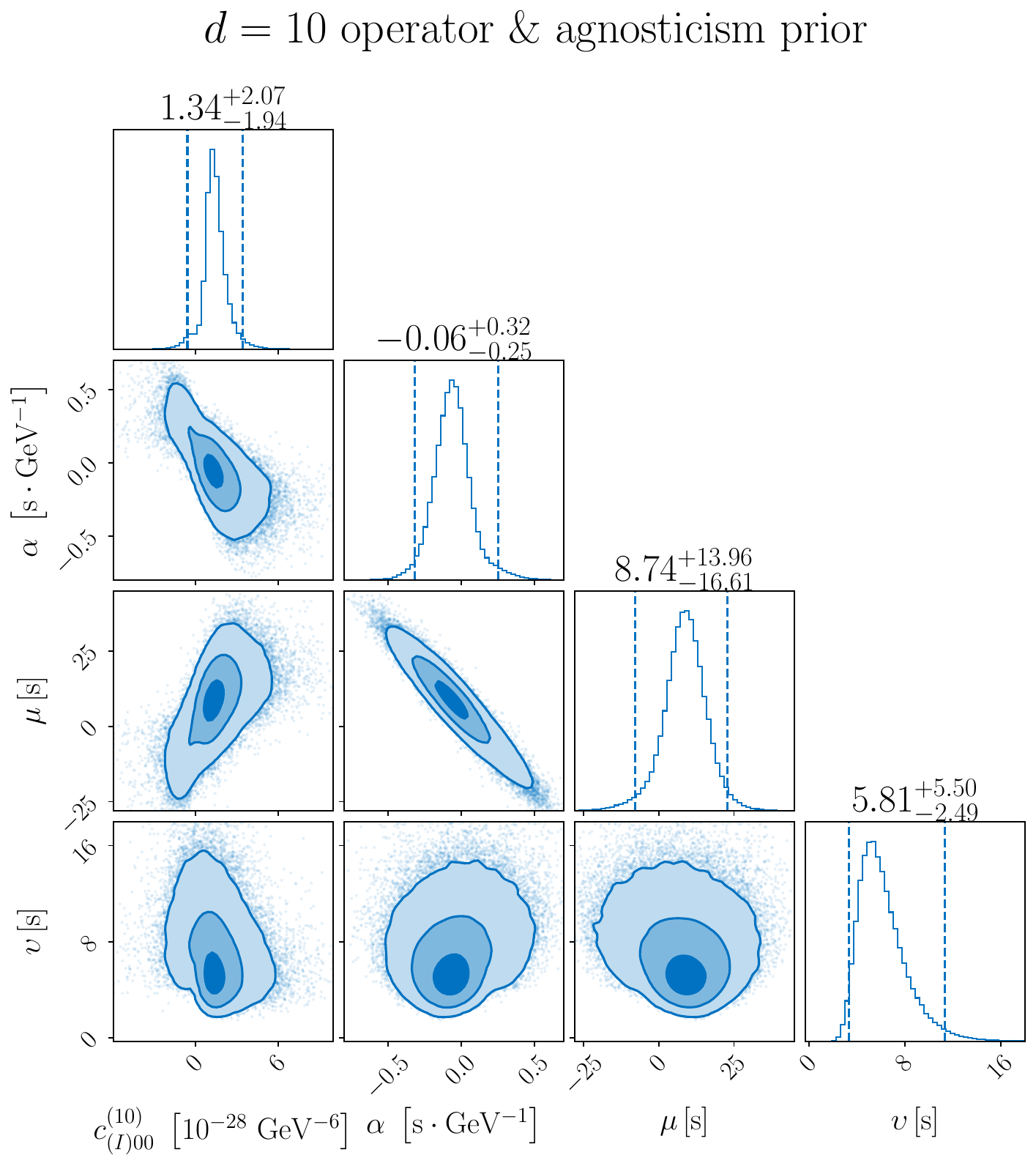} 
    \end{minipage}
    \caption{The results for constraining $d=6$, $d=8$, $d=10$ operators with agnosticism prior are shown in the left, middle, and right panel. The 2D contours represent with different credible levels, denoting the 1$\sigma$, 2$\sigma$, and 3$\sigma$ regions, while the vertical lines indicate the  95\% region for the 1D marginalized posterior distribution. } 
    \label{agnosticism_prior_intrinsic_params}
\end{figure*}

\begin{figure*}[htbp] 
	\centering 
    \begin{minipage}{0.33\textwidth}
        \centering
        \includegraphics[width=0.95\linewidth]{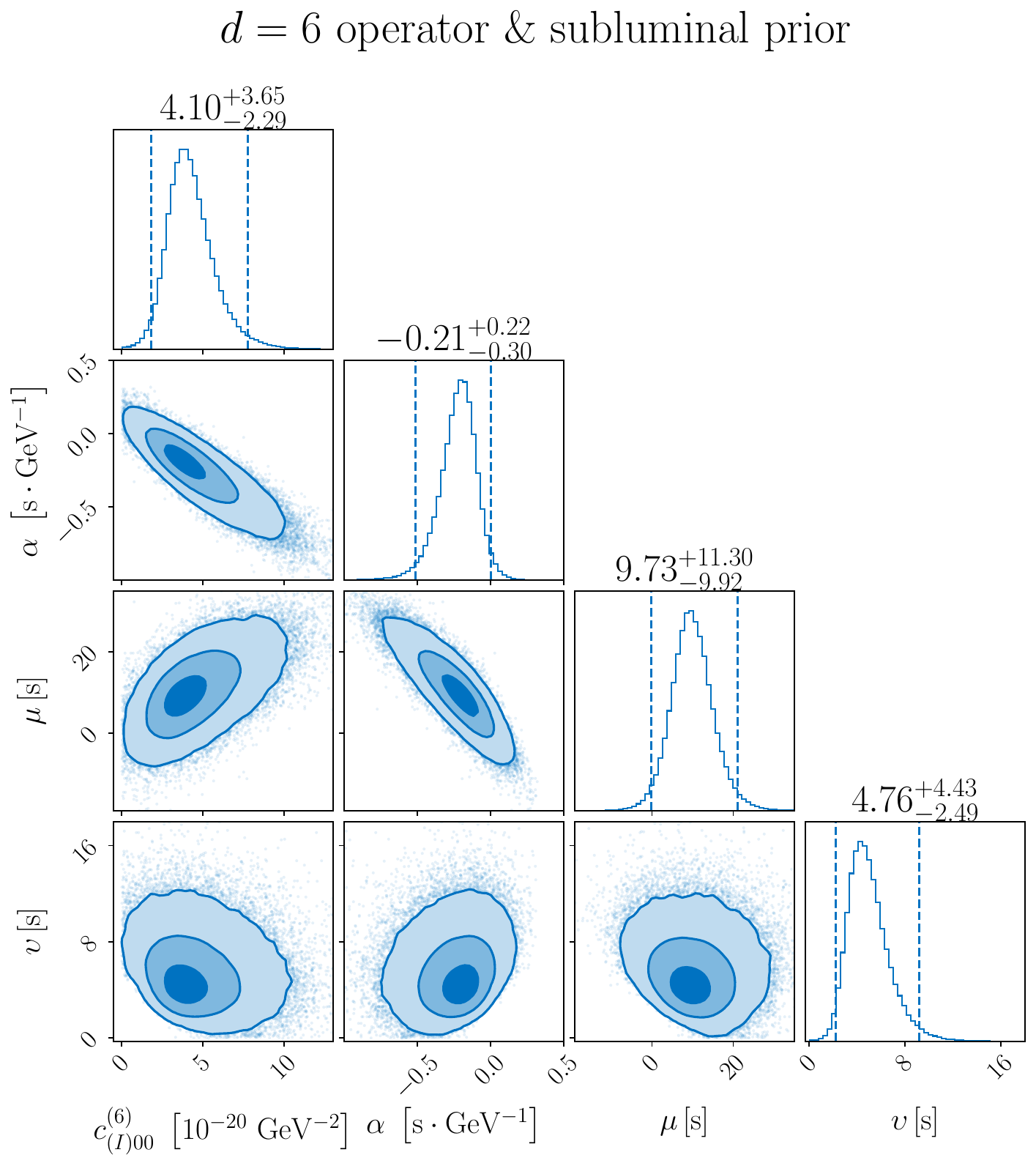}
    \end{minipage}\hfill
    \begin{minipage}{0.33\textwidth}
        \centering
        \includegraphics[width=0.95\linewidth]{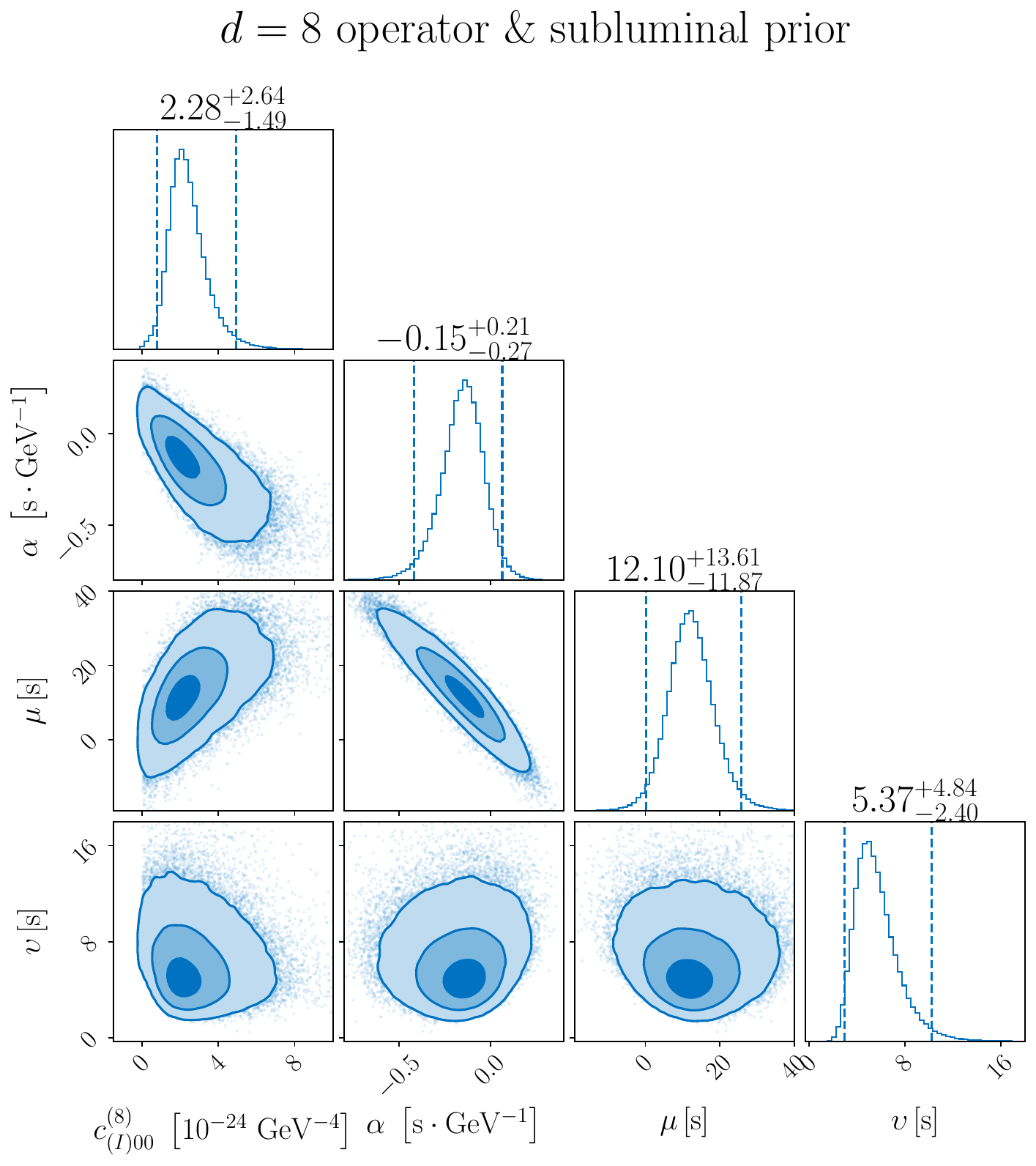} 
    \end{minipage}\hfill
    \begin{minipage}{0.33\textwidth}
        \centering
        \includegraphics[width=0.95\linewidth]{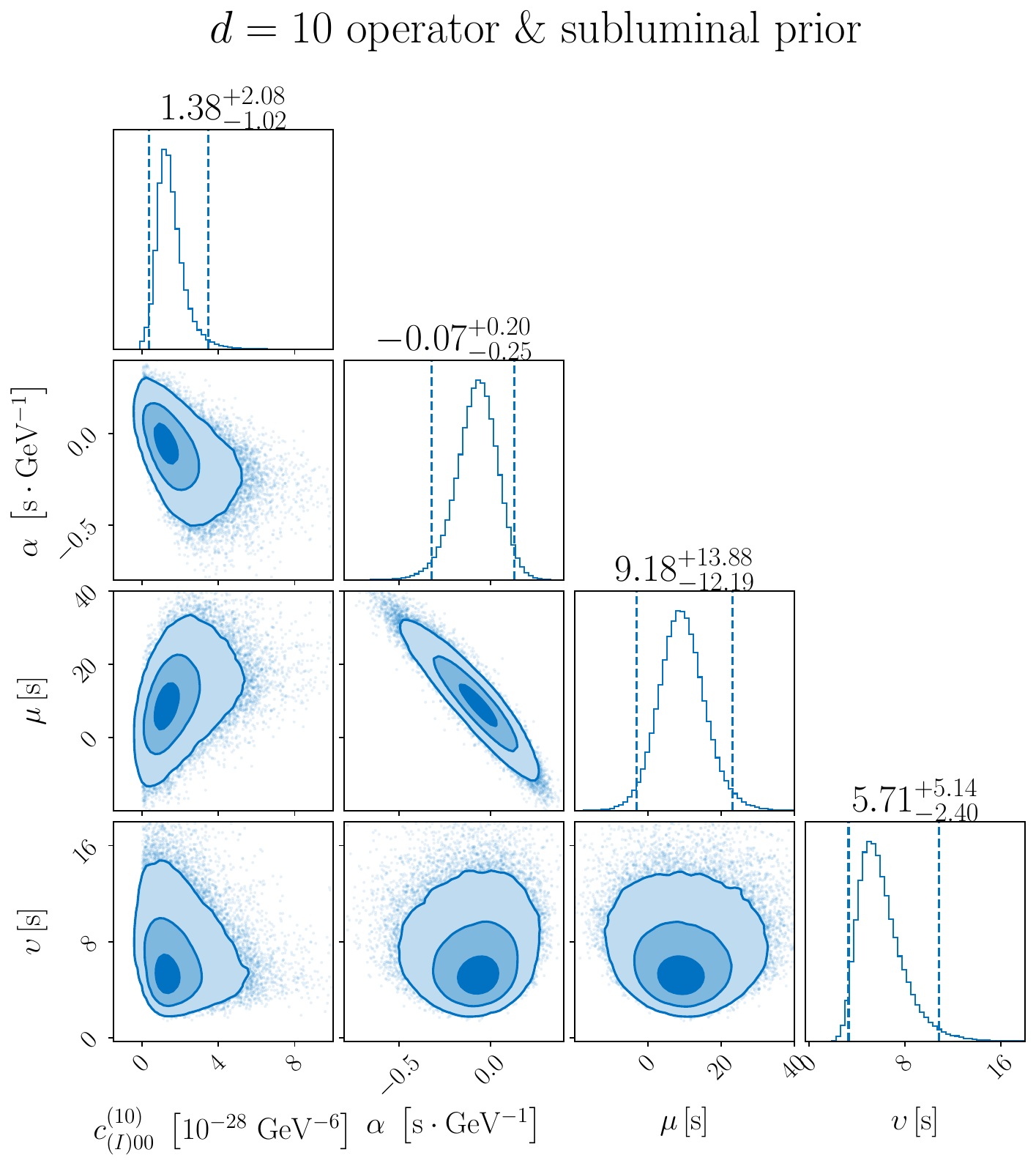} 
    \end{minipage}
    \caption{Same as Fig.~\ref{agnosticism_prior_intrinsic_params} but for subluminal prior.} 
    \label{subluminal_intrinsic_params}
\end{figure*}

\section{Data analysis}\label{DAHS}
In this work, we focus on $c^{(d)}_{(I)jm}$ birefringence-free operator of dimension-6, -8, -10, as defined in Eq.~(\ref{Rel-cF-cI}). For simplicity, we restrict our analysis to the isotopic terms with $j=0$ and $m=0$. As GRB observational data continue to accumulate, particularly the publicly from the Fermi satellite ~\cite{Fermi-LAT:2009pgs, Meegan:2009qu}, the time delay of high-energy GRB photons provides a viable means of constraining the above operators. In this section, we first introduce the GRB data utilized in this work, which consist of 14 GRB photons in the GeV band. We then present in detail the time delay model of GRB photons and the parameter estimation method within the Bayesian framework.

\subsection{GRB data}
Previous studies focused on the spectral lag of GRBs in the keV- and MeV- bands \cite{Wei:2016exb, Wei:2017zuu}, whereas in this work we analyze GRBs with detected high-energy photons in GeV-band ($E_{o}> 10$ GeV). Following Refs.~\cite{Xu:2016zsa, Xu:2016zxi, Song:2024and}, we analyze 14 time lags between GeV-band and the corresponding keV-band photons from eight GRBs, including GRB 080916C~\cite{GRB080916C}, GRB 090510~\cite{GRB090510}, GRB 090902B~\cite{GRB090902B}, GRB 090926A~\cite{GRB090926A}, GRB 100414A~\cite{GRB100414A}, GRB 130427A~\cite{GRB130427A}, GRB 140619B~\cite{GRB140619B}, and GRB 160509A~\cite{GRB160509A}. The GeV-band photons were detected by Fermi-LAT~\cite{Fermi-LAT:2009pgs}, while the keV-band photons were recorded by Fermi-GBM \cite{Meegan:2009qu}. 
\reply{Detailed information, including redshifts, spectral lags and observed photon energies, and the sky localizations of the eight GRBs analyzed in this work are summarized in Table~\ref{datatable}. In line with previous works \citep{Shao:2009bv, Zhang:2014wpb, Xu:2016zxi, Xu:2016zsa}, we adopt low-energy photons (8–260 keV) observed by the Gamma-ray Burst Monitor (GBM) \citep{Meegan:2009qu} as a reference for comparison with high-energy photons in our calculations. Following \citep{Xu:2016zxi, Xu:2016zsa, Liu:2018qrg}, the arrival time of low-energy photons, $t_{l}$, is taken as the first significant peak. For the eight GRBs analyzed here, we use the $t_{l}$ values reported in earlier works \citep{Xu:2016zxi, Xu:2016zsa, Song:2024and}. To account for observational uncertainty, we adopt a positional uncertainty of $\pm 5$~s for the first significant peak in the observer frame \cite{Song:2025myx}.}


\begin{table*}[ht]
\renewcommand{\arraystretch}{1.5}
    \centering
    \caption{\reply{Table of high-energy photon ($>$10 GeV) data from eight GRBs used in this work, including the redshift $z$, sky localizations Right Ascension (RA) and Declination (Dec), the photon energies in the observer frame $E_{\rm h,o}$, the arrival time for high-energy photons $t_{\rm h}$ and for low-energy photons $t_{\rm l}$.}}
    \begin{tabular}{ccccccc}
    \toprule
        GRB name & $z$ & Dec (deg) & RA (deg) & $E_{\rm h,o}$ (GeV) &  $t_{\rm h}$ (s) &  $t_{\rm l}$ (s) \\ 
        \midrule
        GRB 080916C~\cite{GRB080916C} & 4.35 & 119.9 & -56.6 & 12.4 $\pm 1.24$ & 16.55 & 5.98 $\pm 5.00$ \\ 
        GRB 080916C~\cite{GRB080916C} & 4.35 & 119.9 & -56.6 & 27.4 $\pm 2.74$ & 40.51 & 5.98 $\pm 5.00$ \\ 
        GRB 090510~\cite{GRB090510} & 0.903 & 333.4 & -26.8 & 29.9 $\pm 2.99$ & 0.83 & -0.03 $\pm 5.00$ \\
        GRB 090902B~\cite{GRB090902B}  & 1.822 & 265 & 27.3 & 39.9 $\pm 3.99$ & 81.75 & 9.77 $\pm 5.00$ \\ 
        GRB 090902B~\cite{GRB090902B}  & 1.822 & 265 & 27.3 & 11.9 $\pm 1.19$ & 11.67 & 9.77 $\pm 5.00$ \\ 
        GRB 090902B~\cite{GRB090902B}  & 1.822 & 265 & 27.3 & 14.2 $\pm 1.42$ & 14.17 & 9.77 $\pm 5.00$ \\
        GRB 090902B~\cite{GRB090902B}  & 1.822 & 265 & 27.3 & 18.1 $\pm 1.81$ & 26.17 & 9.77 $\pm 5.00$ \\
        GRB 090902B~\cite{GRB090902B}  & 1.822 & 265 & 27.3 & 12.7 $\pm 1.27$ & 42.37 & 9.77 $\pm 5.00$ \\ 
        GRB 090902B~\cite{GRB090902B}  & 1.822 & 265 & 27.3 & 15.4 $\pm 1.54$ & 45.61 & 9.77 $\pm 5.00$ \\
        GRB 090926A~\cite{GRB090926A} & 2.1071 & 353.6 & -66.3 & 19.5 $\pm 1.95$ & 24.84 & 4.32 $\pm 5.00$ \\ 
        GRB 100414A~\cite{GRB100414A} & 1.368 & 191.6 & 8.6 & 29.7 $\pm 2.97$ & 33.37 & 0.29 $\pm 5.00$ \\ 
        GRB 130427A~\cite{GRB130427A} & 0.3399 & 173.2 & 27.7 & 72.6 $\pm 7.26$ & 18.64 & 0.54 $\pm 5.00$ \\ 
        GRB 140619B~\cite{GRB140619B} & 2.67 & 132.7 & -9.7 & 22.7 $\pm 2.27$ & 0.61 & 0.10 $\pm 5.00$ \\ 
        GRB 160509A~\cite{GRB160509A} & 1.17 & 311.3 & 76.1 & 51.9 $\pm 5.19$ & 76.51 & 13.92 $\pm 5.00$ \\
        \bottomrule
    \end{tabular}
    \label{datatable}
\end{table*}

\subsection{Time delay model and Bayesian inference}
In the framework of SME, the observed spectral lags\footnote{In this work, we define the time delay as the arrival time difference between high- and low-energy photons, whereas in Refs.~\cite{Wei:2016exb, Wei:2017zuu}, the time delay is defined as the arrival time difference between low- and high-energy photons.} from GRBs can be attributed to two components: the intrinsic time delay $\Delta t_{\mathrm{in}}$, and LV induced time delay caused by dimension-$d$ operator $\Delta t_{\rm h}^{(d)}$,
\begin{equation}
\label{obs_delay}
\Delta t_{\mathrm{obs}} = t_{\rm h}  - t_{\rm l} = \Delta t_{\mathrm{in}}(1+z) + \Delta t_{\rm LV}^{(d)},
\end{equation}
where $t_{\rm h}$ and $t_{\rm l}$ denote the arrival time of high- and low- energy photons, respectively.  The LV induced time delay associated with dimension-$d$ operator is given by \cite{LVQED2002, Kostelecky:2008be, LVQED2009},
\begin{equation}
    \label{LV_time_delay}
    \Delta t_{\rm LV}^{(d)} = t_{\rm h}^{(d)}  - t_{\rm l}^{(d)}  \simeq (d - 3) (E_{\rm h,o}^{d-4} - E_{\rm l,o}^{d-4})
    \times \int_{0}^{z}\frac{(1+z')^{d-4}\mathrm{d}z'}{H_0\sqrt{\Omega_{m}(1+z')^{3}+\Omega_{\Lambda}}}\sum_{jm}{}_{0}Y_{jm}(\hat{\textbf{n}})c_{(I)jm}^{(d)},
\end{equation}
where $E_{\mathrm{h,o}}^{n}$ and $E_{\rm l,o}^{n}$ denote the energies of high- and low-energy photons, respectively. The parameters $H_0$, $\Omega_{\mathrm{m}}$, and $\Omega_{\Lambda}$ represent the Hubble constant, matter density, and dark energy density of the $\Lambda$CDM model, respectively \cite{Planck:2018vyg}.  
We adopt the standard spherical polar coordinates $(\theta, \phi)$ associated with the direction vector $\hat{\textbf{n}}$, where $\theta = \pi/2 - {\rm Dec}$ and $\phi = {\rm RA}$. 
To accurately analyze GRB observational data, an intrinsic time-delay model must be specified, as first pointed out and implemented in Ref.~\cite{Ellis:2005}. We consider an energy-dependent intrinsic time-delay model, similar to those (albeit in different forms) studied in Refs.~\cite{Ellis:2018lca,Pan:2020zbl,Vardanyan:2022ujc}, and adopt the formulation of Ref.~\cite{Song:2024and},

\begin{equation}
    \Delta t_{\rm{in}} = \Delta t_{\rm in, c} + \alpha E_{\rm h,s},
\end{equation}
where $\Delta t_{\rm in, c}$ is a constant term representing the common intrinsic delay, and $E_{\rm h,s}$ is the source frame energy of high-energy photon, with $\alpha$ being the coefficient. The consistency of this model have been validated by observations of GeV- and TeV-band GRB photons \cite{Song:2024and, Song:2025qej, Song:2025akr}, as well as through Monte Carlo simulations \cite{Song:2025myx}.

Although the full expansion $\sum_{jm}{}_{0}Y_{jm}(\hat{\textbf{n}})c_{(I)jm}^{(d)}$ could in principle be fitted by including infinitely many modes, such an approach is not applicable when combining multiple GRBs, as each event has a different sky-localization unit vector $\hat{\textbf{n}}$. Consequently, when restricting the analysis to a finite number of modes, we retain only the leading-order contribution, namely the monopole term with $j=m=0$ (the isotropic case) for simplicity.

The observed time delay in Eq.~(\ref{obs_delay}), transformed to the source frame, is given by,
\begin{equation}
\frac{\Delta t_{\mathrm{obs}}}{1+z}= \Delta t_{\mathrm{in}} + \frac{\Delta t_{\rm LV}^{(d)}}{1+z} = \Delta t_{\mathrm{in}} + c_{(I)00}^{(d)}S^{(d)}(\hat{\textbf{n}}),
\end{equation}
where the $S^{(d)}(\hat{\textbf{n}})$ is, 
\begin{equation}
\label{s_d}
   S^{(d)}(\hat{\textbf{n}}) =  \frac{(d - 3)}{H_0} \frac{E_{\rm h}^{d-4} - E_{\rm l}^{d-4}}{1+z}
    \times \int_{0}^{z}\frac{(1+z')^{d-4}\mathrm{d}z'}{\sqrt{\Omega_{m}(1+z')^{3}+\Omega_{\Lambda}}}{}_{0}Y_{00}(\hat{\textbf{n}}).
\end{equation}

\hspace{-0.9mm}
With this simplification, once the observational quantities, such as the photon energies $E_{\rm h,o}$ and $E_{\rm l,o}$, redshift $z$ of each GRB,  and cosmological parameters $H_0$, $\Omega_{m}$, and $\Omega_{\Lambda}$ are specified, Eq.~(\ref{s_d}) can be evaluated.

In addition, we employ parameter estimation methods within the Bayesian framework. As proposed in Refs.~\cite{Song:2024and, Song:2025myx}, the likelihood function for the aforementioned time delay model is given by,
\begin{widetext}
    \begin{equation}
    p^{(d)} \propto \exp\left[-\frac{1}{2}\sum_{j=1}^{n}\left(\frac{\left(\frac{\Delta t_{\mathrm{obs},j}}{1+z_{j}} -\mu - \alpha E_{(h,s),j} -c_{(I)00}^{(d)}S^{(d)}(\hat{\textbf{n}})\right)^{2}}{\sigma^2_{y_j}+\upsilon^{2}+ \alpha^2\sigma_{E_{(h,s),j}}^2+(c_{(I)00}^{(d)})^{2}\sigma_{S^{(d)}(\hat{\textbf{n}})}^{2}}+\ln(\sigma^2_{y_j}+\upsilon^{2}+ \alpha^2\sigma_{E_{(h,s),j}}^2+(c_{(I)00}^{(d)})^{2}\sigma_{S^{(d)}(\hat{\textbf{n}})}^{2})\right)\right],
    \end{equation}
\end{widetext}
where the common intrinsic time delay term, $\Delta t_{\rm in,c}$, is assumed to follow a Gaussian distribution, allowing the emission to occur over a finite time interval (denoted by $\upsilon$) \cite{Song:2025myx},
    
\begin{equation}
    p\left(\Delta t_{\rm in,c} \right) \sim \mathcal{N} \left(\mu, \upsilon^2\right).
\end{equation}

Thus, four parameters are needed to be determined from the observed GRB data: the dimension-$d$ operator  $c_{(I)00}^{(d)}$, the coefficient $\alpha$ of the energy-dependent time delay model, the mean value $\mu$ of the common intrinsic time delay term, and its standard deviation $\upsilon$.
The priors for $\alpha$, $\mu$, and $\upsilon$ are assumed to follow uniform (flat) distributions for generality,  
\begin{align}
    \begin{cases}
        p(\alpha) \sim U \left[-50, 50 \right] \ {\rm s \cdot GeV^{-1}}, \\
        p\left(\mu\right) \sim U\left[-50, 50 \right] \ {\rm s}, \\
        p\left(\upsilon\right) \sim U\left[0, 50 \right] \ {\rm s}. \\
    \end{cases}
    \label{priors}
\end{align}
We employ the \texttt{bilby} package \citep{Ashton:2018jfp, Romero-Shaw:2020owr} for our calculations.
\section{Results}

We analyze 14 spectral lags from GeV-band GRB photons to constrain the dimension-6, -8, and -10 operators in the SME framework, namely $c_{(I)00}^{(6)}$, $c_{(I)00}^{(8)}$, and $c_{(I)00}^{(10)}$. In principle, considering these corrections, the speed of light can be either subluminal or superluminal, corresponding to positive or negative values of there operators, respectively. To ensure generality, we first constrain these operators in a theory-agnostic manner using the following priors, 
\begin{align}
    \begin{cases}\label{priors0equal}
        p\left(c_{(I)00}^{(6)}\right) \sim U\left[-50, 50 \right] \times 10^{-20} \ \rm{GeV^{-2}}, \\
        p\left(c_{(I)00}^{(8)}\right) \sim U\left[-50, 50 \right] \times 10^{-24} \ \rm{GeV^{-4}}, \\
        p\left(c_{(I)00}^{(10)}\right) \sim U\left[-50, 50 \right] \times 10^{-28} \ \rm{GeV^{-6}}. \\
    \end{cases}
\end{align}

The results are shown in Fig.~\ref{agnosticism_prior_intrinsic_params} and summarized in Table ~\ref{agnostic_results_table}. The 1$\sigma$ posterior credible interval for all three operators lie entirely within the positive range.  Moreover, the 95\% posterior credible interval of $c_{(I)00}^{(6)}$ and $c_{(I)00}^{(8)}$  also lie in the positive range, with values of $4.11^{+3.64}_{-2.32} \times 10^{-20}~{\rm GeV}^{-2}$ and $2.26^{+2.64}_{-1.70} \times 10^{-24}~{\rm GeV}^{-4}$, respectively. For the operator $c_{(I)00}^{(10)}$, the 95\% posterior credible interval is $1.34^{+2.07}_{-1.94} \times 10^{-28}~{\rm GeV}^{-6}$, where only the lower bound slightly overlaps with the negative range. Therefore, these results suggest that the subluminal case is favored by the observations. 
This is consistent with instability analysis of high-energy $\gamma$-ray photons in the presence of LV, since the would be superluminal photons may be unable to reach us due to its short lifetime by the lepton pair production process: $\gamma\rightarrow l^+\,+\,l^-$, the photon to lepton pair emission.
This rare process is strictly forbidden in the LI case, 
but becomes kinematically allowed once the photon energy exceeds a critical threshold,  
$E_\gamma\ge\frac{2m_l}{\sqrt{-\varsigma^0}}$ \cite{Threshold03}, where $m_l$ denotes the lepton mass ({\it e.g.}, $m_e$),
and $-\varsigma^0\equiv p\cdot\hat{c}_F\cdot p/\omega^2= v_g-1$ represents 
the deviation of the photon group velocity from the LI value.
Since most GRB photons originate from extra-galactic distances, the very existence of ultrahigh-energy GRB photons themselves (the absence of the superluminal instability issue) imposes extremely severe constraints on the LV parameter through $-\varsigma^0$ \cite{SteckerGlashow01}.
On the other hand, the subluminal photon can also cause instability issues for ultrahigh-energy leptons through the vacuum Cherenkov-like radiation $l^-\rightarrow\, l^-+\gamma$.
However, it is precisely the subluminal constraint that we aim to probe through photon arrival time-delay, and 
our results (at least $|-\varsigma^0|<10^{-18}$ 
already improve upon the previous stringent bound of 
$|-\varsigma^0|<9\times10^{-16}$ \cite{KlinkSchreck08} by one order of magnitude.
The latter bounds was obtained using the existence of $10\sim100$TeV electrons as inferred from the observed TeV $\gamma$-ray photons produced via the inverse Compton process \cite{TeV-gamma1997,SteckerGlashow01}. 
Meanwhile, the central values of the coefficients $\alpha$ are negative for all three cases, consistent with the results of Ref.~\cite{Wei:2017zuu}, which analyzed spectral lags of keV- and MeV-band GRB photons. 

In the following, we focus on constraining the dimension-6, -8, and -10 operators for the subluminal case. Flat priors are also assumed for all operators,
\begin{align}
    \begin{cases}\label{priors-positive}
        p\left(c_{(I)00}^{(6)}\right) \sim U\left[0, 50 \right] \times 10^{-20} \ \rm{GeV^{-2}}, \\
        p\left(c_{(I)00}^{(8)}\right) \sim U\left[0, 50 \right] \times 10^{-24} \ \rm{GeV^{-4}}, \\
        p\left(c_{(I)00}^{(10)}\right) \sim U\left[0, 50 \right] \times 10^{-28} \ \rm{GeV^{-6}}. \\
    \end{cases}
\end{align}

The results are shown in Fig.~\ref{subluminal_intrinsic_params} and summarized in Table ~\ref{results_table}. These findings are consistent with the aforementioned theory-agnostic test, supporting the robustness of the Bayesian parameter estimation method. We obtain the 95\% posterior credible  intervals of $c_{(I)00}^{(6)}$, $c_{(I)00}^{(8)}$, and $c_{(I)00}^{(10)}$ as $4.10^{+3.65}_{-2.29} \times 10^{-20} ~ {\rm GeV}^{-2}$, $2.28^{+2.64}_{-1.49} \times 10^{-24} ~ {\rm GeV}^{-4}$, and $1.38^{+2.08}_{-1.02} \times 10^{-28} ~ {\rm GeV}^{-6}$, respectively, representing improvements of 6, 12, and 19 orders of magnitude for the dimension-6, -8, and -10 operators compared to Wei et al. \cite{Wei:2017zuu}. These improvement are mainly due to the inclusion of GeV-band photons in this work, compared to the keV- and MeV-band photons considered by Wei {\it et al.} \cite{Wei:2017zuu}, as shown in Eq.~(\ref{LV_time_delay}). This also explains why we obtain much tighter constraints for higher-dimension operators, whereas Wei {\it et al.} \cite{Wei:2017zuu} achieved tighter constraints for lower-dimension operators.


It is noteworthy that all the above results are tighter than the strongest constraints listed in the annually updated table of LV limits \cite{DataTable2025}. 
For example, previous studies constrained the dimension-6, -8, -10 operators as $|c_{(I)00}^{(6)}|=4.25^{+1.60}_{-1.63}\times10^{-15}~\mathrm{GeV}^{-2}$~\cite{Wei:2022zqi}, $|c_{(I)00}^{(8)}|=5.38^{+1.84}_{-1.84}\times10^{-12}~\mathrm{GeV}^{-4}$ 
~\cite{Wei:2022zqi}, and $|c_{(I)00}^{(10)}|<3.5\times10^{-9}~\mathrm{GeV}^{-6}$~\cite{Wei:2017zuu}, respectively. In comparison, our results represent improvements of 5, 12, and 19 orders of magnitude over the above constraints, as shown in Table~\ref{results_table}.


\begin{table*}[h]
\renewcommand{\arraystretch}{1.5}
\setlength{\tabcolsep}{2\tabcolsep}
  \centering
  \caption{Table of constraining dimension-6, -8, and -10 SME operators in the theory-agnostic way. The error bars denotes the 95\% region of 1D marginalized posterior distribution.}
    \begin{tabular}{ccccc}
    \toprule
    Case & $c_{(I)00}^{(d)} ~ ({\rm GeV}^{-(d-4)})$ & $\alpha ~( \rm{s} \cdot {\rm GeV}^{-1}) $ & $\mu ~({\rm s})$ & $\upsilon ~({\rm s})$  \\
    \midrule
    $d$ =6 & $4.11^{+3.64}_{-2.32} \times 10^{-20}$ & $-0.21^{+0.22}_{-0.30}$ & $9.74^{+11.30}_{-10.01}$ & $4.78^{+4.53}_{-2.51}$  \\
    $d$ =8 & $2.26^{+2.64}_{-1.70} \times 10^{-24}$ & $-0.15^{+0.24}_{-0.27}$  & $11.94^{+13.67}_{-12.95}$ & $5.39^{+5.03}_{-2.42}$ \\
    $d$ =10 & $1.34^{+2.07}_{-1.94} \times 10^{-28}$ & $-0.06^{0.32}_{-0.25}$ & $8.74^{+13.96}_{-16.61}$ & $5.81^{+5.50}_{-2.49}$  \\
    \bottomrule
    \end{tabular}%
  \label{agnostic_results_table}%
\end{table*}

\begin{table*}[h]
\renewcommand{\arraystretch}{1.5}
\setlength{\tabcolsep}{2\tabcolsep}
  \centering
  \caption{Table of constraining the subluminal case of dimension-6, -8, and -10 SME operators. The error bars denotes the 95\% region of 1D marginalized posterior distribution. The third column presents the results  obtained from the analysis of keV- and MeV-band GRB photons in Wei {\it et al}.~\cite{Wei:2017zuu}.}
    \begin{tabular}{cccccc}
    \toprule
    Case & $c_{(I)00}^{(d)} ~ ({\rm GeV}^{-(d-4)})$ & $c_{(I)00}^{(d)}$ from Wei {\it et al}~\cite{Wei:2017zuu} $({\rm GeV}^{-(d-4)})$ & $\alpha ~( \rm{s} \cdot {\rm GeV}^{-1}) $ & $\mu ~({\rm s})$ & $\upsilon ~({\rm s})$  \\
    \midrule
    $d$ =6 & $4.10^{+3.65}_{-2.29} \times 10^{-20}$ & $< 1.2 \times 10^{-14} $ & $-0.21^{+0.22}_{-0.30}$ & $9.73^{+11.30}_{-9.92}$ & $4.76^{+4.43}_{-2.49}$  \\
    $d$ =8 & $2.28^{+2.64}_{-1.49} \times 10^{-24}$ & $< 7.2 \times 10^{-12}$ & $-0.15^{+0.21}_{-0.27}$  & $12.10^{+13.61}_{-11.87}$ & $5.73^{+4.84}_{-2.40}$ \\
    $d$ =10 & $1.38^{+2.08}_{-1.02} \times 10^{-28}$ & $< 3.5 \times 10^{-9}$ & $-0.07^{0.20}_{-0.25}$ & $9.18^{+13.88}_{-12.19}$ & $5.71^{+5.14}_{-2.40}$  \\
    \bottomrule
    \end{tabular}%
  \label{results_table}%
\end{table*}

\section{Discussion}\label{results}
In this work, we focus exclusively on the birefringent-free subset of 
LV
coefficients in the photon sector within the framework of the 
SME. 
The underlying reason is twofold:
(1). vacuum birefringence is already subject to extremely stringent constraints from astrophysical polarization measurements \cite{Shao:2011uc,Astro-K1, Astro-K2, Astro-K3} and CMB observations \cite{CMBLV-N,CMBLV-K1, CMBLV-K2, CMBLV-K3};
and (2). within the 
EFT
framework, especially the SME, it is essential to explore the birefringence-free subset of LV photon operators, 
since model-independent and birefringence-free dispersion relations are frequently exployed in studies of photon time-of-flight delays.
However, It should be noted that in EFT-based constructions there appears to be no exactly birefringence-free subset valid to all orders in the LV coefficients.
Using 14 recorded spectral lags of GeV photons from GRB observations
and assuming flat priori probability distributions, we place very stringent constraints on
the isotropic, power-counting nonrenormalizable LV coefficients $c_{(I)00}^{(n)}$ for 
$n=6,8,10$ within the SME framework. 

The constraint procedure is first carried out in a theory-agnostic manner.
Since the resulting $95\%$ posterior credible intervals almost exclude any superluminal behavior, we then refine the analysis by restricting the flat priors to the subluminal regime, namely, to positive intervals.
The results obtained in a theory-agnostic manner are summarized in Table \ref{agnostic_results_table},
while those obtained under the prior assumption of subluminal propagation are presented in Table \ref{results_table}.
The two sets of results 
are mutually consistent.
By combining the theory-agnostic and subluminal-prior
estimates from Tables \ref{agnostic_results_table} and \ref{results_table}, respectively, we obtain the following more conservative bounds on the isotropic birefringence-free LV parameters, 
$|c_{(I)00}^{(6)}|\le7.75 \times 10^{-20} ~ {\rm GeV}^{-2}$, $|c_{(I)00}^{(8)}|\le4.92 \times 10^{-24} ~ {\rm GeV}^{-4}$, and $|c_{(I)00}^{(10)}|\le3.46 \times 10^{-28} ~ {\rm GeV}^{-6}$.
The preference for the subluminal scenario is also consistent with earlier stability  analyses \cite{SteckerGlashow01,Threshold03},
which relied on superluminal $\gamma$ photon decay and electron vacuum Cherenkov radiation \cite{Marco2015,Altschul07}.
Although the results clearly favor the subluminal scenario, 
we emphasize that they are based on only 14 spectral lags of GeV photons,
and may change when much large data sets become available in the future.

Even with such limited GeV $\gamma$ photon events, our constraints can still  surpass previous constraints with credible intervals by at least five orders of magnitude, and are at least comparable to the most stringent limits reported in \cite{DataTable2025}. The tighter bounds arise primarily from the use of GeV-band $\gamma$-ray photons, rather than lower-energy photons \cite{Wei:2017zuu}, 
although this comes at the cost of reduced statistical significance due to the limited number of GeV-band $\gamma$-ray photon events. 
However, as more high-energy GRB events are accumulated, the statistical significance can be gradually improved.
We expect that high-energy astrophysical messengers may finally help to address 
some long-standing issues, such as whether Lorentz symmetry is exact or not?

\emph{\textbf{Acknowledgements}.---}
We thank A. Kosteleck\'y and the anonymous referee for valuable comments and suggestions.
This work is supported by National Natural Science Foundation of China
under Grant No. 12335006, 
and by High-performance Computing Platform of Peking University.

\bibliography{scibib}

\end{document}